\documentclass[]{article}

\usepackage{savesym}
\usepackage{graphicx}

\usepackage{authblk}
\usepackage{lineno}

\usepackage{xcolor}

\savesymbol{tablenum}
\usepackage{siunitx}

\newcommand\unit[3][]{\SI[#1]{#2}{#3}}
\restoresymbol{SIX}{tablenum}

\usepackage[version=3]{mhchem}
\usepackage{makecell}

\usepackage{geometry}
\usepackage[utf8]{inputenc}       
\usepackage[T1]{fontenc}         

\usepackage{amsmath}
\usepackage{amssymb}
\usepackage{mathrsfs}

\usepackage{lineno}

\usepackage{tikz}
\usetikzlibrary{plotmarks,patterns,calc,decorations.markings}

\usepackage{booktabs}

\usepackage{lipsum}

\usepackage[symbol]{footmisc}

\catcode`\@=11 
\def\parenbar{\mathpalette\p@renb@r}
\def\p@renb@r#1#2{\vbox{%
\ifx#1\scriptscriptstyle \dimen@.7em\dimen@ii.2em\else
\ifx#1\scriptstyle \dimen@.8em\dimen@ii.25em\else
\dimen@1em\dimen@ii.4em\fi\fi \offinterlineskip
\ialign{\hfill##\hfill\cr
\vbox{\hrule width\dimen@ii}\cr
\noalign{\vskip-.3ex}%
\hbox to\dimen@{$\mathchar300\hfil\mathchar301$}\cr
\noalign{\vskip-.3ex}%
$#1#2$\cr}}}
\catcode`\@=12 
\def\nuan{\parenbar{\nu}\kern-0.4ex}

\newlength{\smfigwidth}
\newlength{\figwidth}
\newlength{\captwidth}
\usepackage[pdflatex]{trimclip}

\usepackage{hyperref}
\usepackage[all]{hypcap}

\title{ANTARES and IceCube Combined Search for Neutrino Point-like and Extended Sources in the Southern Sky}

\date{}

\begin{document}

\author[ ]{ANTARES Collaboration\footnote{Email: antares.spokesperson@in2p3.fr}:}\setcounter{authors}{0}

\author[1,2]{A.~Albert}
\author[3]{M.~Andr\'e}
\author[4]{M.~Anghinolfi}
\author[5]{G.~Anton}
\author[6]{M.~Ardid}
\author[7]{J.-J.~Aubert}
\author[8]{J.~Aublin}
\author[8]{B.~Baret}
\author[9]{S.~Basa}
\author[10]{B.~Belhorma}
% ...

\affil[1]{\scriptsize{Universit\'e de Strasbourg, CNRS,  IPHC UMR 7178, F-67000 Strasbourg, France}}
\affil[2]{\scriptsize{Universit\'e de Haute Alsace, F-68200 Mulhouse, France}}
\affil[3]{\scriptsize{Technical University of Catalonia, Laboratory of Applied Bioacoustics, Rambla Exposici\'o, 08800 Vilanova i la Geltr\'u, Barcelona, Spain}}
\affil[4]{\scriptsize{INFN - Sezione di Genova, Via Dodecaneso 33, 16146 Genova, Italy}}
\affil[5]{\scriptsize{Friedrich-Alexander-Universit\"at Erlangen-N\"urnberg, Erlangen Centre for Astroparticle Physics, Erwin-Rommel-Str. 1, 91058 Erlangen, Germany}}
\affil[6]{\scriptsize{Institut d'Investigaci\'o per a la Gesti\'o Integrada de les Zones Costaneres (IGIC) - Universitat Polit\`ecnica de Val\`encia. C/  Paranimf 1, 46730 Gandia, Spain}}
\affil[7]{\scriptsize{Aix Marseille Univ, CNRS/IN2P3, CPPM, Marseille, France}}
\affil[8]{\scriptsize{APC, Univ Paris Diderot, CNRS/IN2P3, CEA/Irfu, Obs de Paris, Sorbonne Paris Cit\'e, France}}
\affil[9]{\scriptsize{Aix Marseille Univ, CNRS, CNES, LAM, Marseille, France }}
\affil[10]{\scriptsize{National Center for Energy Sciences and Nuclear Techniques, B.P.1382, R. P.10001 Rabat, Morocco}}
% ...

\setcounter{authors}{0}
\author[ ]{\ \\ \ \\}\setcounter{authors}{0}
\author[ ]{IceCube Collaboration\footnote{Email: analysis@icecube.wisc.edu}:}\setcounter{authors}{0}

\author[61]{M. G. Aartsen}
\author[100]{M. Ackermann}
\author[61]{J. Adams}
\author[57]{J. A. Aguilar}
\author[65]{M. Ahlers}
\author[91]{M. Ahrens}
\author[71]{C. Alispach}
\author[82]{K. Andeen}
\author[97]{T. Anderson}
\author[57]{I. Ansseau}

% ...
\affil[57]{\scriptsize{Universit{\'e} Libre de Bruxelles, Science Faculty CP230, B-1050 Brussels, Belgium}}
% ...
\affil[61]{\scriptsize{Dept. of Physics and Astronomy, University of Canterbury, Private Bag 4800, Christchurch, New Zealand}}
% ...
\affil[65]{\scriptsize{Niels Bohr Institute, University of Copenhagen, DK-2100 Copenhagen, Denmark}}
% ...
\affil[71]{\scriptsize{D{\'e}partement de physique nucl{\'e}aire et corpusculaire, Universit{\'e} de Gen{\`e}ve, CH-1211 Gen{\`e}ve, Switzerland}}
% ...
\affil[82]{\scriptsize{Department of Physics, Marquette University, Milwaukee, WI, 53201, USA}}
% ...
\affil[91]{\scriptsize{Oskar Klein Centre and Dept. of Physics, Stockholm University, SE-10691 Stockholm, Sweden}}
% ...
\affil[97]{\scriptsize{Dept. of Physics, Pennsylvania State University, University Park, PA 16802, USA}}
% ...
\affil[100]{\scriptsize{DESY, D-15738 Zeuthen, Germany}}
% ...

\maketitle


\begin{abstract}
A search for point-like and extended sources of cosmic neutrinos using data collected by the ANTARES and IceCube neutrino telescopes is presented. The data set consists of all the track-like and shower-like events pointing in the direction of the Southern Sky included in the nine-year ANTARES point-source analysis, combined with the through-going track-like events used in the seven-year IceCube point-source search. The advantageous field of view of ANTARES and the large size of IceCube are exploited to improve the sensitivity in the Southern Sky by a factor $\sim$2 compared to both individual analyses. In this work, the Southern Sky is scanned for possible excesses of spatial clustering, and the positions of preselected candidate sources are investigated. In addition, special focus is given to the region around the Galactic Centre, whereby a dedicated search at the location of SgrA* is performed, and to the location of the supernova remnant RXJ~1713.7-3946. No significant evidence for cosmic neutrino sources is found and upper limits on the flux from the various searches are presented.
\end{abstract}

\section{Introduction}
\label{intro}

Neutrinos are stable, neutral, and weakly interacting particles and in contrast to cosmic rays, they are not deflected by magnetic fields. Differently from high-energy photons, neutrinos are effectively not absorbed while traveling through cosmological distances and can escape from dense astrophysical environments. These qualities make them ideal cosmic messengers as they point back to their production sites. Several classes of astrophysical objects, like supernova remnants, pulsar wind nebulae and active galactic nuclei have been indicated as promising high-energy neutrino source candidates~\cite{Sources1, Sources2, Sources3, Sources4, Sources5}. Neutrinos are expected to be produced through the decay of charged mesons, results of hadronic interactions of accelerated protons with matter or radiation in the surroundings of the acceleration sites. \newline \indent 
Neutrino astronomy has recently entered an exciting period with the discovery of an isotropic high-energy cosmic neutrino flux reported by the IceCube Collaboration~\cite{IC3years, IC4yproc, IC6yproc, IC-VHE-29, IC-VHE-36}, followed by the first evidence of neutrino emission from an astrophysical source, the blazar TXS~0506+056 \cite{ICTXS, ICmultimessenger}. These observations represent a major breakthrough in the field, thus further investigations are strongly motivated. Indeed, the origin of most of the observed neutrino flux remains unknown. The neutrino flux of TXS~0506+056 can only account for less than 1\% of the total observed astrophysical flux~\cite{ICTXS}. Moreover, recent searches for neutrino emission from the directions of blazars from the 2nd Fermi-LAT AGN catalogue performed by the IceCube Collaboration indicated that blazars contribute less than about 40\% - 80\% (30\%) to the total observed neutrino flux assuming an unbroken power-law spectrum $\Phi(E_{\nu}) \propto E_{\nu}^{-\gamma}$ with $\gamma = 2.0$~\cite{ICTXS} ($\gamma =2.5$~\cite{ICblazars}). The region around TXS~0506+056 was studied also by the ANTARES Collaboration using data collected from 2007 to 2017~\cite{ANTAREStxs}. The standard time-integrated method fits 1.03 signal events, which corresponds to a p-value of 3.4\% (not considering trial corrections).

In this work, the point-source data samples of the ANTARES~\cite{antaresdetector} and IceCube~\cite{icecubedetector, icecubedetector2} neutrino telescopes collected during nine~\cite{lastPSant} and seven years~\cite{lastPSic}, respectively, are combined to perform various searches for point-like and extended sources of neutrinos in the Southern Sky. This work supersedes a previous combined analysis using a smaller data sample of five and three years of ANTARES and IceCube data, respectively~\cite{firstcombined}. 

The two telescopes complement each other thanks to their different characteristics, in particular the larger instrumented volume of IceCube and the privileged view of the Southern Sky with a reduced muon background for neutrino energies below 100 TeV of ANTARES. Exploiting these different characteristics allows for a significant gain in sensitivity for searches in the Southern Sky. 

The paper is organised as follows: a brief description of the ANTARES and IceCube neutrino telescopes is given in Section \ref{detectors}. In Section \ref{samples}, the samples employed in the searches are described. The analysis method and the expected performances are discussed in Section \ref{method}, while the performed searches and corresponding results are presented in Section \ref{searches}. In Section \ref{Conc}, conclusions are drawn.

\section{ANTARES and IceCube neutrino telescopes}
\label{detectors}

The ANTARES and IceCube neutrino telescopes rely on the same principle for detecting cosmic neutrinos. A three-dimensional array of photomultiplier tubes (PMTs) inside a transparent medium -- water and ice, respectively --, collects the Cherenkov photons induced by the passage of relativistic charged particles. The charged particles are produced in neutrino interactions with the target medium, inside or near the instrumented volume. The information provided by the number of detected Cherenkov photons and their arrival times is used to infer the neutrino interaction topology, direction and energy.

The ANTARES telescope~\cite{antaresdetector} is located in the Mediterranean Sea, 40 km South of Toulon (France), at a depth of about 2400 m. It was completed in 2008, with the first lines operating since 2006. The detector comprises a three-dimensional array of 885 optical modules (OMs), each one housing a 10” PMT, facing \unit{45}{\degree} downward in order to optimise the detection of Cherenkov photons from upgoing charged particles. The PMTs are distributed over 12 vertical lines with a length of 350 m, and with an inter-line separation between 60 and 75 m, instrumenting a total volume of $\sim$0.01$\ \textrm{km}^3$.

The IceCube telescope~\cite{icecubedetector, icecubedetector2} is a cubic-kilometer-sized detector located at the South Pole, between 1450 and 2450 m below the surface of the Antarctic ice. A total of 5160 digital optical modules (DOMs), each consisting of a pressure-resistant sphere that houses electronics, calibration LEDs, and a 10” PMT facing downward, are attached to 86 vertical strings, with a mean distance between strings of $\sim$125 m. The construction of the IceCube detector began in 2005 and was finished six years later. During the construction, data were collected with partial configurations of the detector, commonly indicated by ICXY, with XY denoting the number of active strings. 

Two main event topologies can be identified in the ANTARES and IceCube telescopes: tracks and showers. Charged current (CC) interactions of muon neutrinos and antineutrinos produce a relativistic muon that can travel large distances through the medium. Cherenkov light is emitted along the muon path leaving a track-like signature in the detector. Shower-like events are induced by neutral current (NC) interactions, and by CC interactions of electron and tau neutrinos and antineutrinos. They are characterised by an almost spherically symmetric light emission around the shower maximum. The longer lever arm of the track topology allows for a better reconstruction of the particle direction and therefore for a better median angular resolution, making tracks more suited than showers to search for point-like sources. On the other hand, a better reconstruction of the particle energy is achieved for showers, as the topology allows for a calorimetric measurement. 

Common backgrounds in both detectors are atmospheric muons and neutrinos originating from cosmic ray interactions in Earth’s atmosphere. Events from the Southern Sky correspond to down-going events for IceCube. In this case, atmospheric muons represent the bulk of the detected events before selection, outnumbering the atmospheric neutrinos by a factor from $10^4$ up to $10^6$ depending on the direction. In contrast, ANTARES' detected events are predominantly atmospheric neutrinos because the Earth acts as a shield for atmospheric muons.

\section{Data samples}
\label{samples}

All track-like and shower-like events from the Southern Sky which were employed in the nine-year ANTARES point-source analysis~\cite{lastPSant}, combined with the through-going track-like events, i.e.\ tracks induced by muons traversing the detector, included in the seven-year IceCube point-source search~\cite{lastPSic} are used in this analysis. The ANTARES data were collected between early 2007 and the end of 2015. The IceCube data were taken from 2008 to 2015, with the detector operating either in partial (samples IC40, IC59, IC79) or in full (samples IC86-2011, IC86-2012-2015) configuration. 

The ANTARES events were selected by applying cuts on the zenith angle, the angular error estimate and parameters describing the quality of the reconstruction. In the case of the shower events, a cut was also applied on the interaction vertex, required to be located within a fiducial volume slightly larger than the instrumented volume. 
A detailed description of these cuts can be found in~\cite{lastPSant}. 
The selection criteria were optimised to minimise the neutrino flux needed for a $5\sigma$ discovery of a point-like source emitting with a $E_{\nu}^{-2.0}$ spectrum. 
The selection yielded a total of 7622 (180) tracks (showers) in the whole sky, with 5807 (102) of these events in the Southern Sky. 
A median angular resolution better than $0.4^{\circ}$ is achieved for the selected tracks for energies above 100 TeV, and $\sim$3$^{\circ}$ for the selected showers for energies between 1 TeV and 0.5 PeV. 

The IceCube selection of through-going tracks in the Southern Sky was based on multivariate selection techniques (boosted decision tree, BDT) to discriminate signal from the large background due to down-going atmospheric muons~\cite{lastPSic}. 
The BDT made use of parameters connected to the event quality, track topology, energy deposited along the track, and light-arrival time of photons at the DOMs.
The final event selection was also optimised to yield the best sensitivity and discovery potential for an $E_{\nu}^{-2.0}$ spectrum. This procedure selects only very high-energy events ($E_{\nu} \gtrsim 100$ TeV), yielding a total number of 325 969 events in the five samples. The selected track events are reconstructed with a median angular resolution better than $0.4^{\circ}$ for energies above 1 PeV. 

A summary of the data sets in terms of livetime and number of selected events in each sample and for each detector is given in Table~\ref{tabSamples}. 
\begin{table}[!ht]	
\centering
\caption{Overview over the seven data samples of ANTARES and IceCube employed in the analysis. Only Southern-sky events (numbers of events reported in the last column) have been selected for the present analysis. \newline}
\label{tabSamples}
\begin{tabular}{cccc}
\hline \hline
ANTARES sample & index $j$ & Livetime $T$ [days] & Number of events  \\\hline 
Tracks & 1 & 2415 & 5807 \\
Showers & 2 & 2415 & 102 \\\hline \hline
IceCube sample & index $j$ & Livetime $T$ [days] & Number of events  \\\hline 
IC40 & 3 & 376 & 22779 \\
IC59 & 4 & 348 & 64257 \\
IC79 & 5 & 316 & 44771 \\
IC86-2011 & 6 & 333 & 74931 \\
IC86-2012-2015 & 7 & 1058 & 119231\\\hline
\end{tabular}
\end{table}
As a consequence of the different layouts, locations of the telescopes and selection techniques in the Southern Sky, each sample has a different efficiency for detecting events from potential sources. The relative contribution $C^j(\delta, \Phi)$ for each sample $j$, defined as the ratio of the expected mean number of signal events for the given sample to that for all samples, $C^j = N^j/\sum_{i=1}^7 N^i$, depends on the expected flux from the source $\Phi$ and declination $\delta$. For each ANTARES sample $j \in [1,2]$ given in Table~\ref{tabSamples}, the expected mean number of signal events, $N^j$, is obtained as~\cite{lastPSic}:

\begin{align} \label{eq:exp}
N^j = \sum_{f \in \{ \mu, e, \tau \} } T^j \int d\Omega \int dE_{\nu}\ A^{j,\  \nu_{f} + \bar{\nu}_{f}}_{\rm eff} (E_{\nu}, \Omega)\ \Phi_{\nu_{f} + \bar{\nu}_{f}}(E_{\nu}, \Omega),
\end{align} 

\noindent where the contribution of each neutrino flavour $f$ to the track and shower channels is considered. $T^j$ is the livetime of the sample $j$ reported in Table~\ref{tabSamples}, $\Omega$ is the solid angle, $E_{\nu}$ is the neutrino energy, $A^{j, \ \nu_{f} + \bar{\nu}_{f}}_{\rm eff}$ is the detector effective area, and $\Phi_{\nu_{f} + \bar{\nu}_{f}}$ is the expected flux from the source. The expected mean number of signal events for each IceCube sample $j \in [3,7]$ given in Table~\ref{tabSamples} is calculated using Equation \ref{eq:exp}, including only the contribution of the muon flavour.
Unless otherwise stated, an unbroken power law neutrino flux is used in the analysis:
\begin{align} \label{eq:Unbrokenspec}
\Phi_{ \nu_{f} + \bar{\nu}_{f}} = \Phi_0 \ \left( \frac{E_\nu}{1\,{\rm GeV}} \right)^{-\gamma},
\end{align} 
with $\Phi_0$ being the one-flavour neutrino flux normalization. Equipartition at Earth of the three neutrino flavours is assumed.

\begin{figure*}[ht!]
\centering
\includegraphics[width=0.48\linewidth]{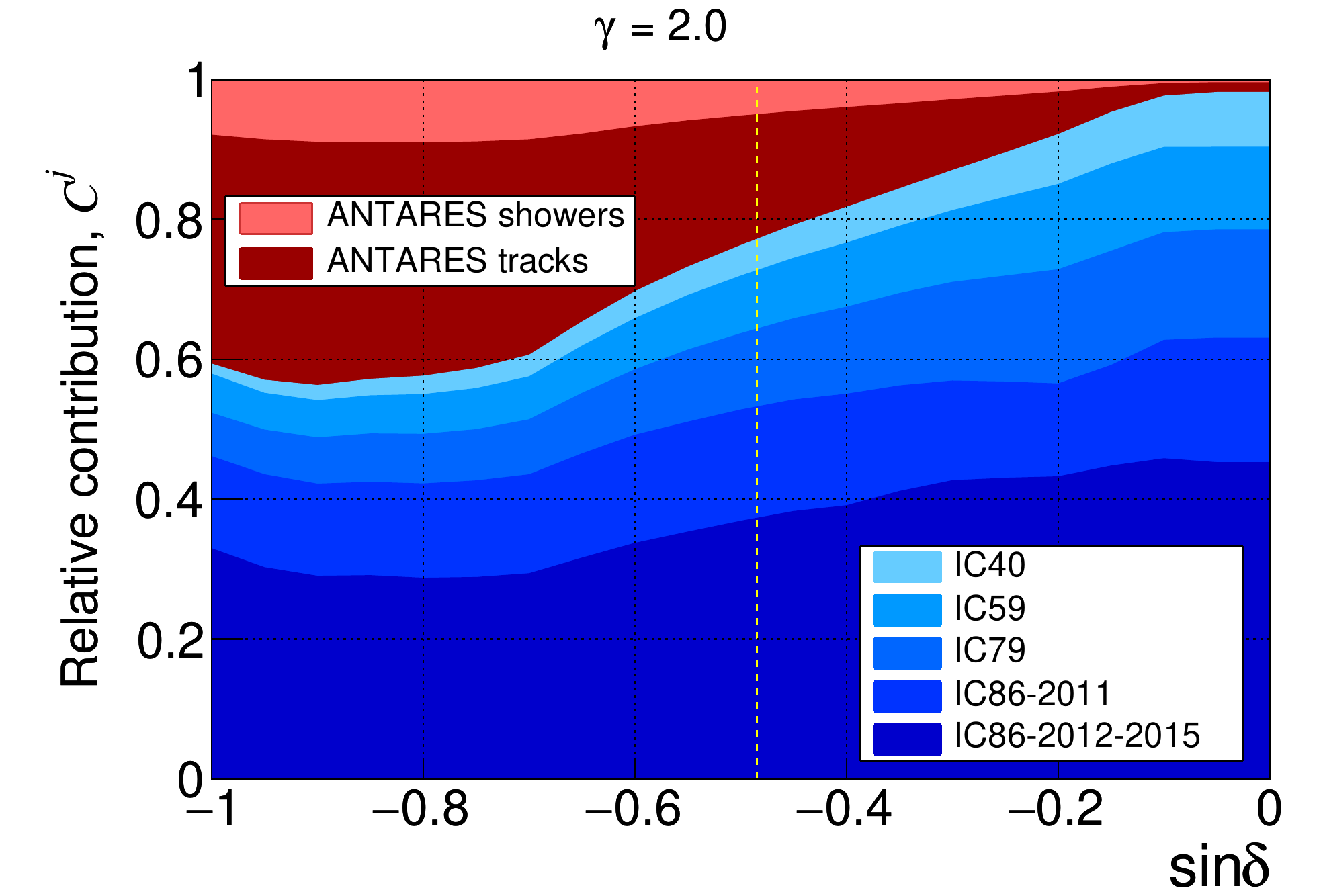}
\includegraphics[width=0.48\linewidth]{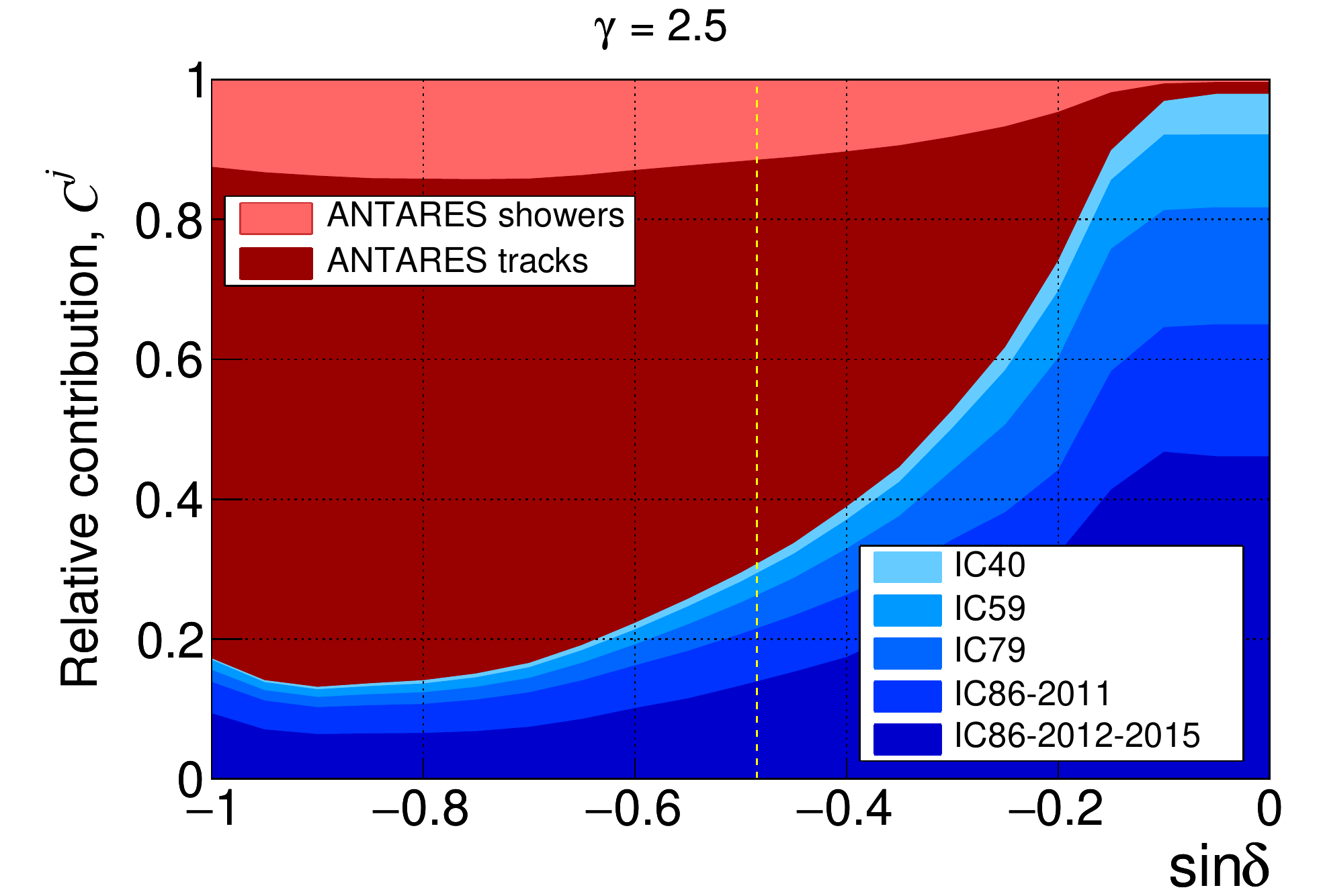}      
\caption{Relative contribution of each sample as a function of the source declination for an unbroken $E_{\nu}^{-\gamma}$ spectrum, with $\gamma = 2.0$ (left) and $\gamma = 2.5$ (right). The contribution of the ANTARES (IceCube) samples is represented by different shades of red (blue). The vertical dashed line marks the declination of the Galactic Centre.}
\label{fig:relAcc}
\end{figure*}

Figure \ref{fig:relAcc} shows the relative contribution of each sample as a function of the source declination for the unbroken $E_{\nu}^{-\gamma}$ spectrum for two values of the spectral index, $\gamma = 2.0$ and $\gamma = 2.5$. The two spectral indices account for the value predicted by the Fermi acceleration mechanism ($\gamma = 2.0$) and for the softer best-fit spectral indices of the isotropic flux of high-energy cosmic neutrinos measured by the IceCube Collaboration (the chosen value for the soft spectral index lies between $\gamma = 2.92$ obtained in~\cite{IC6yproc} and $\gamma = 2.28$ obtained in~\cite{diffuse10}).
For an $E_{\nu}^{-2.0}$ spectrum all samples contribute significantly to most of the Southern Sky. For the softer spectrum $E_{\nu}^{-2.5}$, the contribution of high-energy neutrinos is lower, and therefore, the relative contribution of the ANTARES sample increases.

\section{Search method}
\label{method}

An unbinned likelihood maximization is used to identify clusters of cosmic neutrinos from point-like and extended sources over the randomly distributed atmospheric background.
The likelihood describes the data in terms of signal and background probability density functions (PDFs) and is defined as:

\begin{align} \label{eq:lik}
   L(n_{\rm s}, \gamma, \alpha, \delta) = \prod_{j = 1}^{7} \prod_{i = 1}^{ N^{j}} \Big[ \frac{n_{\rm s}^{j}}{N^{j}} {S}^{ j}_{i}(\gamma, \alpha, \delta) + \Big( 1 - \frac{n_{\rm s}^{j}}{N^{j}} \Big) {B}^{j}_{i} \Big],
\end{align} 

\noindent where $n_{\rm s}$ and $\gamma$ are respectively the unknown total number of signal events and signal spectral index, and $\alpha$ and $\delta$ are the unknown equatorial coordinates of the source. $S^j_i$ and $B^j_{i}$ are the values of the signal and background PDFs for the event $i$ in the sample $j$. $N^{j}$ is the total number of data events in sample $j$, while $n_{\rm s}^{j}$ is the unknown number of signal events in sample $j$, related to $n_{\rm s}$ through the relative contribution of the given sample, $n_{\rm s}^j = n_{\rm s} \cdot \ C^j(\delta, \Phi)$.

The signal and background PDFs are given by the product of a directional and an energy-dependent term. The same definition of the ANTARES and IceCube PDFs used in the respective individual point-source analyses~\cite{lastPSant, lastPSic} is employed in this search. For the IceCube samples, the spatial PDF is given by a 2-dimensional Gaussian, $P_{\rm{space}}^{\rm{IC}} = \rm{exp}(-\Delta\Psi_{\mathnormal i}^2/2\sigma_{\mathnormal i}^2)/(2\pi\sigma_{\mathnormal i}^2)$, with $\Delta\Psi_i$ being the angular distance of the event from the source and $\sigma_i$ being the angular error estimate of the event. When searching for spatially extended sources, the value of $\sigma_i$ is replaced with $\sigma_{\rm eff, \mathnormal i} = \sqrt{\sigma_i^2 + \sigma_{\rm s}^2}$, where $\sigma_{\rm s}$ is the extension of the source assuming a Gaussian profile. For the ANTARES samples, a parameterisation of the point-spread function (PSF) is used as spatial signal PDF.
It is defined as the PDF to reconstruct an event at a given angular distance from its original direction due to reconstruction uncertainties, and it is derived from Monte Carlo simulations. The original direction of the events is given by the location of the source in the case of a point-source hypothesis. For extended sources, the PSF is built assuming that the original direction of the events is distributed according to a Gaussian profile around the centre of the source location with standard deviation given by the assumed source extension $\sigma_{\rm s}$.

During the likelihood maximisation the number of signal events $n_{\rm s}$ and the signal spectral index $\gamma$ are free parameters. Moreover, the position of the source is either kept fixed or fitted within specific limits depending on the type of search (see Section \ref{searches}).

The test statistic, $Q$, is defined as: 

\begin{equation}
     Q = 2 (\log  L(\hat{n}_{\rm s}, \hat{\gamma},  \hat{\alpha}, \hat{\delta}) - \log  L(n_{\rm s} = 0)),
    \label{eq:teststat}
\end{equation}

\noindent where $\hat{n}_{\rm s}$, $\hat{\gamma}$, $\hat{\alpha}$ and $\hat{\delta}$ are the best-fit values that maximise the likelihood. In order to estimate the significance of any observation, the observed $Q$ is compared to the test statistic distribution obtained with background-only pseudo-experiments (PEs) -- pseudo-data sets of data randomised in time to eliminate any local clustering due to potential sources. The fraction of background-like pseudo-experiments with a value of $Q$ larger than the observed $Q$-value gives the significance (p-value) of the observation.

The free parameters can vary over a certain parameter space. The spectral index $\gamma$ can range between 1.0 and 4.0, as these are the limits of reasonable spectral assumptions for astrophysical particle acceleration mechanisms. The lower limit of $n_{\rm s}$ is set to 0.001 in order to have a proper estimation of the median sensitivity, i.e.\ the median expected 90\% C.L. upper limit on the flux normalization in case of pure background. Indeed, if the lower boundary, $n_{\rm s}^{\rm{min}}$, is set to $n_{\rm s}^{\rm{min}} = 0$, a test statistic $Q = 0$ is obtained in more than 50\% of the PEs, leading to an over-estimation of the median 90\% upper limit. By setting $n_{\rm s}^{\rm{min}}$ slightly above 0, the test statistic $Q$ gets negative values for under-fluctuation of the signal. This makes it possible to properly calculate the median of the background $Q$-distribution. 



\begin{figure*}[ht!]
\centering
\includegraphics[width=0.48\linewidth]{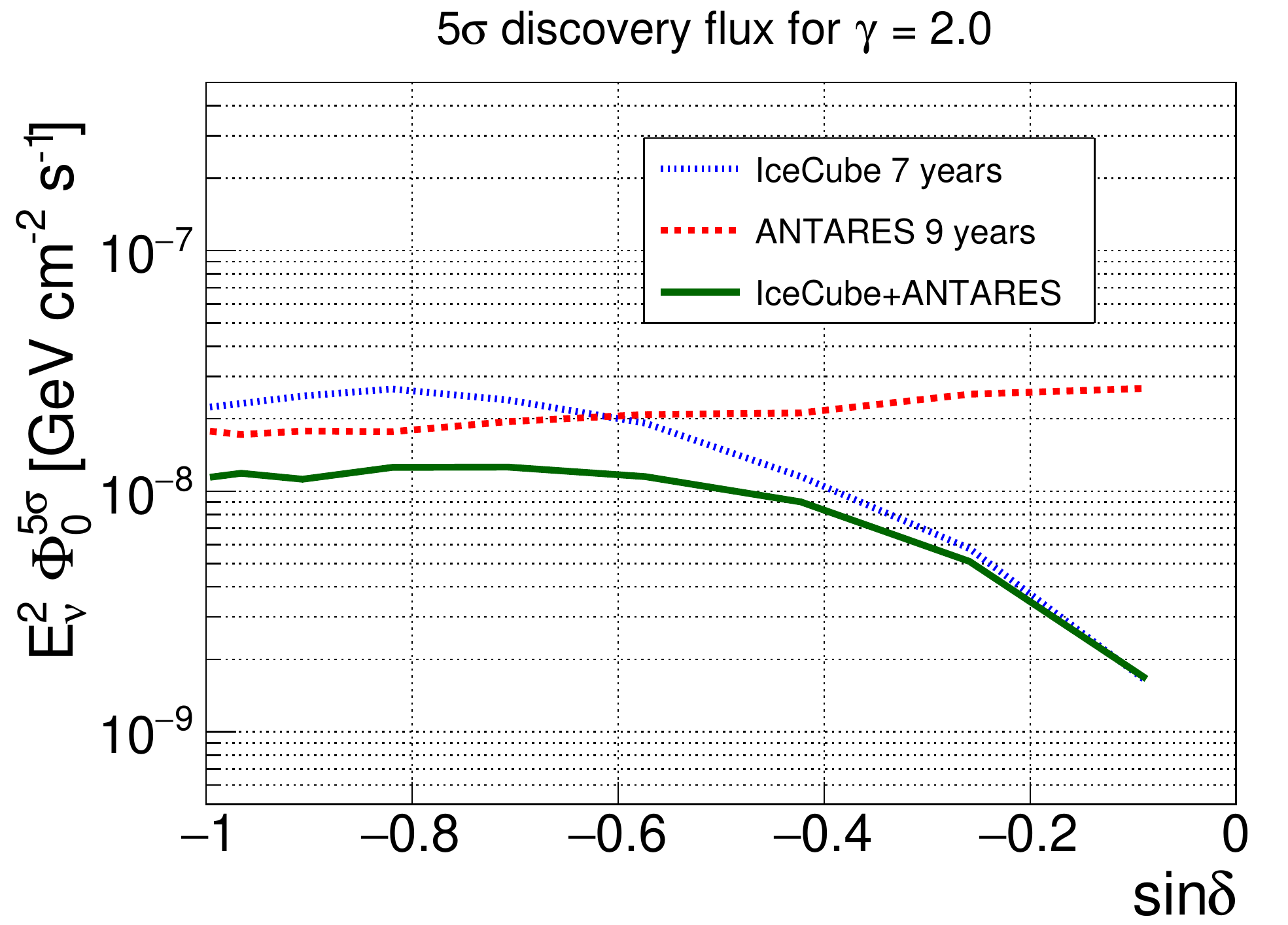}
\includegraphics[width=0.48\linewidth]{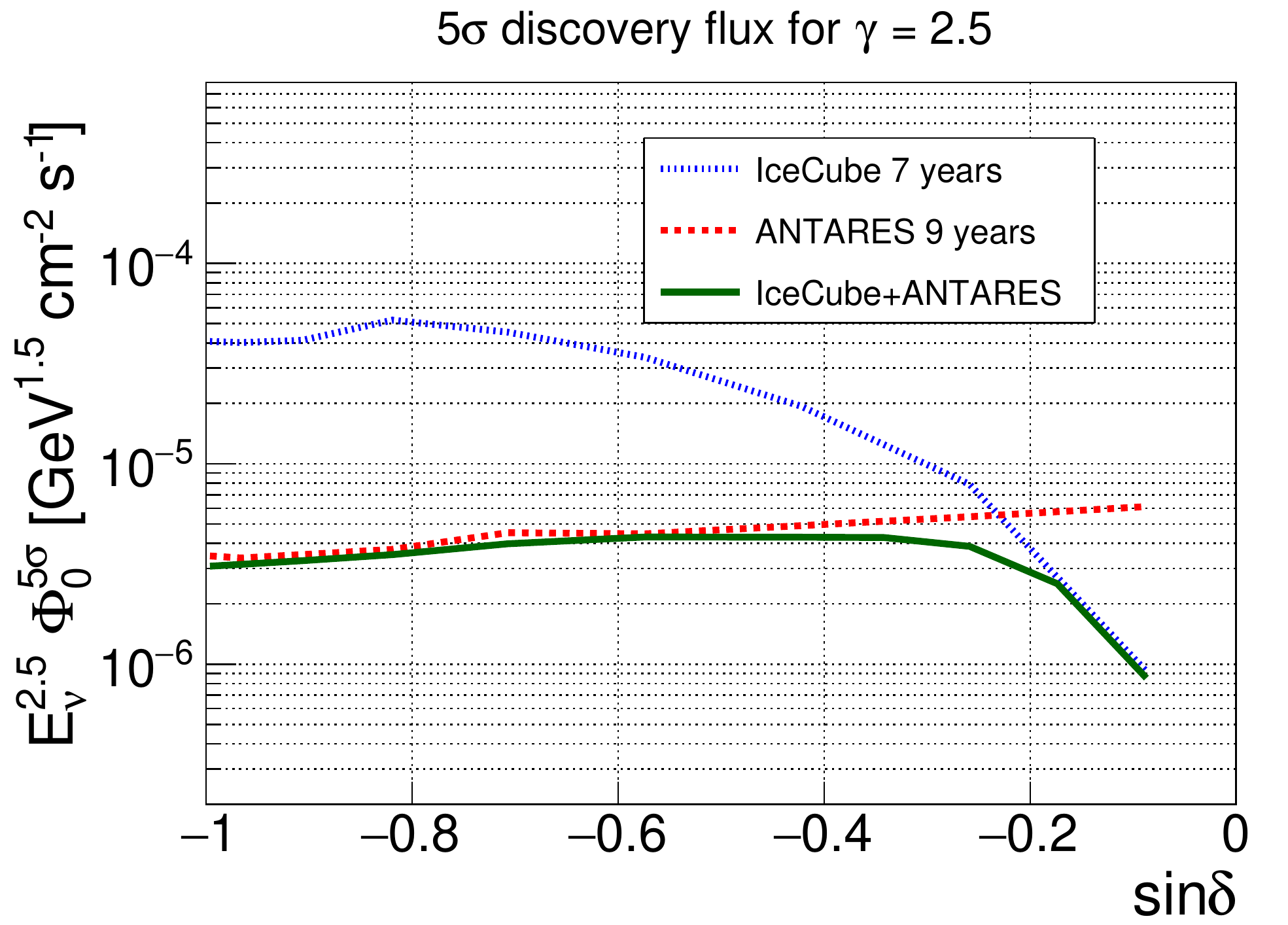}      

\caption{Point-source $5\sigma$ discovery fluxes for an unbroken $E_{\nu}^{-\gamma}$ neutrino spectrum, with $\gamma = 2.0$ (left) and $\gamma = 2.5$ (right). The green line indicates the results for the combined search. Blue and red curves show the results for the individual IceCube and ANTARES analyses, respectively.}
\label{fig:DiscSens}
\end{figure*}

To estimate the potential of the combined search to discover a neutrino source, the $5\sigma$ discovery flux, i.e.\ the neutrino flux needed for a $5\sigma$ discovery in 50\% of the trials, is calculated for an $E_{\nu}^{-\gamma}$ neutrino spectrum, with $\gamma$ equal to 2.0 and 2.5, as a function of the declination. The results are shown in Figure \ref{fig:DiscSens} in comparison to the discovery potentials from the individual IceCube and ANTARES analyses (sensitivities are shown in Figure \ref{fig:LimitsCL}). The discovery flux improves by a factor $\sim$2 in different regions of the Southern Sky, depending on the energy spectrum of the source, compared to the individual IceCube and ANTARES analyses. This result is consistent with the findings of the previous combined analysis~\cite{firstcombined}. For an $E_{\nu}^{-2.0}$ spectrum, the largest improvement is achieved in a region of the sky that is centred approximately at the declination of the Galactic Centre (sin$\delta \sim -0.5$).

\section{Searches and Results}
\label{searches}

Five types of searches for point-like and extended sources are performed in this analysis. In the first two searches, a scan of the full Southern Sky and of a restricted region around the Galactic Centre (GC) are carried out. In the third one, the directions of a pre-defined list of known sources which are potential neutrino emitters are investigated. Finally, we perform two dedicated searches at the locations of two promising neutrino source candidates, the super-massive black hole (SMBH) Sagittarius A*, and the shell-type supernova remnant (SNR) RXJ~1713.7-3946.

\subsection{Southern-sky search and Galactic Centre region search}
\label{SSS}

The most significant clustering with respect to the expected background is searched for at any position in a pre-defined region of the sky. To find the largest deviation from background expectation, the scanned region is divided into a grid with boxes of $1^{\circ} \times 1^{\circ}$ in right ascension and declination. In each box, the unbinned likelihood maximisation is performed, with the source position being an additional parameter, free to vary inside the $1^{\circ} \times 1^{\circ}$ boundaries. For each box, the best-fit values of the number of signal events, $\hat{n}_{\rm s}$, spectral index,  $\hat{\gamma}$, source equatorial coordinates, $\hat{\delta}$ and $\hat{\alpha}$, and the test statistic $Q$ are obtained. By comparing the $Q$-value observed at the location of the fitted cluster with the $Q$-distribution obtained from background-only PEs at the corresponding declination, the pre-trial p-value is calculated. The direction with the smallest p-value identifies the most significant cluster of each search. As many directions in the sky are observed, a trial correction must be taken into account when estimating the significance of the observation. To this purpose, the pre-trial p-value is compared to the distribution of the smallest p-values found when performing the same analysis on many background-only PEs.
The fraction of background-like PEs with a p-value smaller than the observed pre-trial p-value gives the trial-corrected significance (post-trial p-value) of the observation.
 
\begin{figure*}[ht!]
\centering
\includegraphics[width=0.7\linewidth]{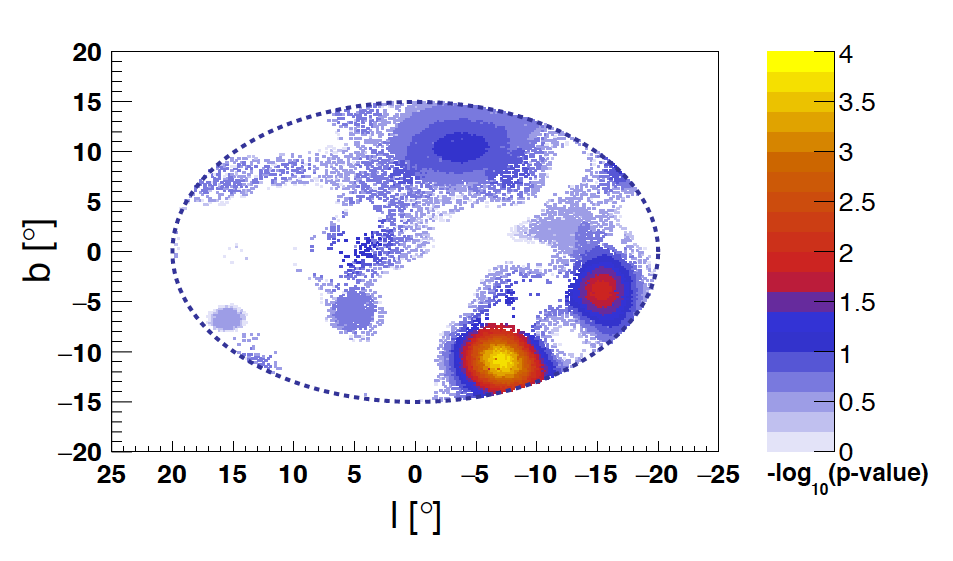}
\caption{Sky map in galactic coordinates of the pre-trial p-values obtained in the Galactic Centre search for the extended source hypothesis with $\sigma_{\rm s} = 2.0\si{\degree}$. The dashed line depicts the boundary of the search area.}
\label{fig:MapGC}
\end{figure*}

In the first search, the scanned region is defined by the whole Southern Sky. Given the large number of probed directions, the significance of weak sources is reduced due to a high trial correction. Motivated by the high concentration of candidate sources and gas around the GC and the recent observation of a possible Pevatron presence close to the GC by the HESS Collaboration~\cite{HESSPeV}, the second search is concentrated around the GC. The examined region (depicted in Figure \ref{fig:MapGC}) is defined by an ellipse centred in the origin of the galactic coordinate system $(\alpha, \delta) = (266.40^{\circ}, -28.94^{\circ})$.  

\begin{figure*}[ht!]
\centering
\includegraphics[width=1.\linewidth]{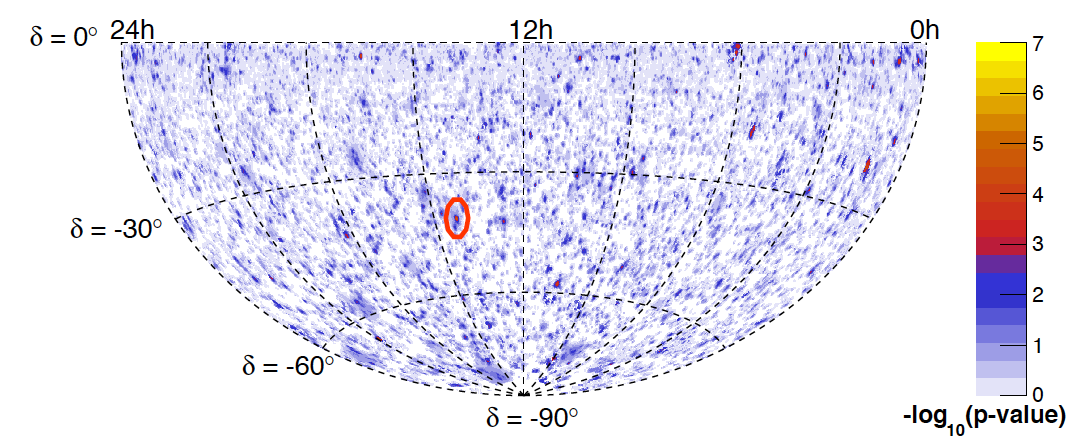}
\caption{Sky map in equatorial coordinates of the pre-trial p-values obtained in the Southern-sky search for the point-like source hypothesis. The red contour indicates the location of the most significant cluster.}
\label{fig:MapFSS}
\end{figure*}

The results are presented in Tables~\ref{tab:FSStable} and~\ref{tab:GCtable}. In both cases, searches for emission regions assumed as point-like ($\sigma_{\rm s} = 0.0^\circ$) or extended ($\sigma_{\rm s} = 0.5^\circ$, $1.0^\circ$, $2.0^\circ$) are performed. For each search and source-extension hypothesis, the best-fit values of the parameters and the p-value of the most significant cluster are reported.
The largest excess above background in the whole Southern Sky is found at equatorial coordinates ($\hat{\alpha}, \hat{\delta}) = (213.2\si{\degree}, -40.8\si{\degree})$, for a point-like source hypothesis, with best-fit $\hat{n}_{\rm s}= 5.7$ and $\hat{\gamma} = 2.5$. A pre-trial p-value of $1.3 \times 10^{-5}$ is obtained for this cluster. The corresponding post-trial significance, obtained by correcting the pre-trial significance for the trials incurred by testing multiple locations, is 18\% (0.9$\sigma$ in the one-sided sigma convention). Figure \ref{fig:MapFSS} depicts the pre-trial p-values for all the investigated directions for a point-like source hypothesis. The position of the most significant cluster is also indicated.

\begin{figure*}[ht!]
\centering
\includegraphics[width=0.5\linewidth]{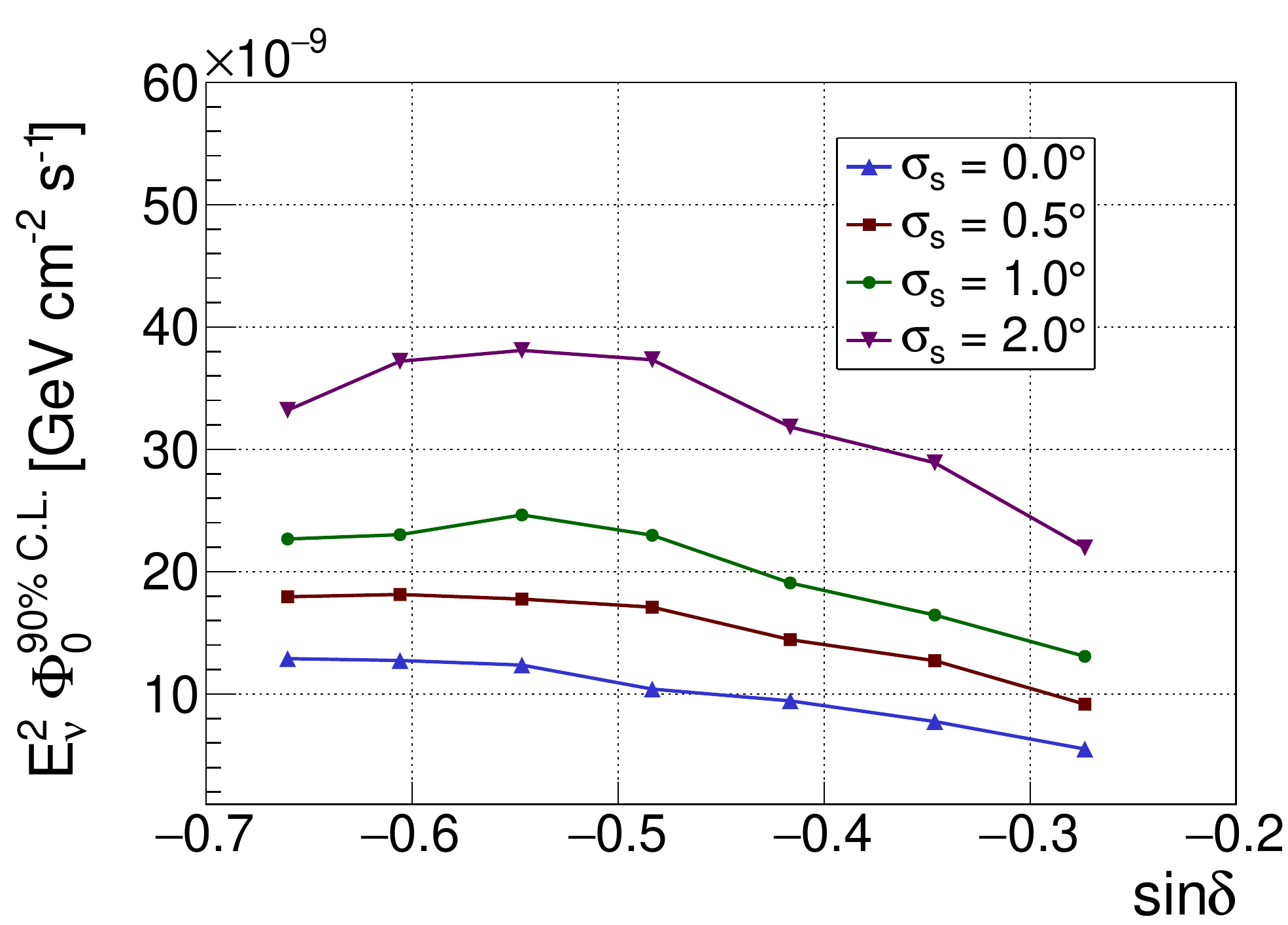}
\caption{90\% C.L. upper limits on the one-flavour neutrino flux normalization of the Galactic Centre region search assuming an $E_{\nu}^{-2.0}$ spectrum for different source extensions $\sigma_{\rm s}$.}
\label{fig:LimitsGC}
\end{figure*}

The most significant result of the search restricted to the Galactic Centre region is observed for an extended source hypothesis ($\sigma_{\rm s} = 2.0\si{\degree}$) at equatorial coordinates ($\hat{\alpha}, \hat{\delta}) = (274.1\si{\degree}, -40.1\si{\degree})$, and galactic coordinates ($\hat{l}, \hat{b}) = (-6.7\si{\degree}, -11.0\si{\degree})$. The values of the best-fit $\hat{n}_{\rm s}$ and $\hat{\gamma}$ are 20.3 and 3.0, respectively. The significance of the hotspot already corrected for looking at multiple directions (post-trial significance) is 3\% (1.9$\sigma$ in the one-sided sigma convention). 
Figure \ref{fig:MapGC} shows the pre-trial p-values for the investigated directions in the Galactic Centre region for an extended source hypothesis with $\sigma_{\rm s} = 2.0\si{\degree}$.
The declination-dependent 90\% C.L. upper limits on the one-flavour neutrino flux normalization of this search are shown in Figure \ref{fig:LimitsGC} for different source extensions. In this analysis, the Neyman method~\cite{neyman} is used to derive sensitivities and limits.

\setlength{\tabcolsep}{.3em}
\begin{table*}[ht!]
    \label{tab:FSStable}
    \caption{List of the most significant clusters found when performing the Southern-sky search for different source-extension hypotheses. Reported are the source extension $\sigma_{\rm s}$, the best-fit parameters (number of signal events, $\hat{n}_{\rm s}$, spectral index, $\hat{\gamma}$,  declination, $\hat{\delta}$, right ascension, $\hat{\alpha}$), and the pre-trial and post-trial p-values.\medskip}
    \resizebox{\linewidth}{!}{
        \begin{minipage}{0.8\textwidth} 
        \centering
 {\def\arraystretch{1.2}
    \footnotesize
    \begin{tabular}{c|c|c|c|c|c|c}
   \hline
            $\sigma_{\rm s} [\si{\degree}]$ & $\hat{n}_{\rm s}$ & $\hat{\gamma}$ & $\hat{\delta} [\si{\degree}]$   &    $\hat{\alpha} [\si{\degree}]$ & pre-trial p-value & post-trial p-value  \\
\hline

0.0 & 5.7 & 2.5 & -40.8 & 213.2 & $1.3 \times 10^{-5}$  & 0.18 \\ 
0.5 & 10.5 & 3.9 & -22.5 & 18.5 & $3.4 \times 10^{-5}$ & 0.31 \\ 
1.0 & 11.6 & 3.8 & -21.9 & 18.4 & $8.9 \times 10^{-5}$ & 0.44 \\ 
2.0 & 20.3 & 3.0 & -40.1 & 274.1 & $2.2 \times 10^{-4}$ & 0.47 \\ 

\end{tabular}
      }
      \end{minipage}
    }
\end{table*}

\setlength{\tabcolsep}{.3em}
\begin{table*}[ht!]
    \label{tab:GCtable}
    \caption{List of the most significant clusters found when performing the search in the Galactic Centre region for different source-extension hypotheses. Reported are the source extension $\sigma_{\rm s}$, the best-fit parameters (number of signal events, $\hat{n}_{\rm s}$, spectral index, $\hat{\gamma}$,  declination, $\hat{\delta}$, right ascension, $\hat{\alpha}$), and the pre-trial and post-trial p-values.    \medskip}
    \resizebox{\linewidth}{!}{
        \begin{minipage}{0.8\textwidth} 
        \centering
 {\def\arraystretch{1.2}
    \footnotesize
    \begin{tabular}{c|c|c|c|c|c|c}
   \hline
            $\sigma_{\rm s} [\si{\degree}]$ & $\hat{n}_{\rm s}$ & $\hat{\gamma}$ & $\hat{\delta} [\si{\degree}]$   &    $\hat{\alpha} [\si{\degree}]$ & pre-trial p-value & post-trial p-value  \\
\hline

0.0 & 6.8 & 2.8 & -42.3 & 273.0 & $7.3 \times 10^{-4}$  & 0.40 \\ 
0.5 & 8.4 & 2.8 & -42.0 & 273.1 & $5.2 \times 10^{-4}$ & 0.19 \\ 
1.0 & 12.1 & 2.9 & -41.8 & 274.1 & $6.9 \times 10^{-4}$ & 0.15 \\ 
2.0 & 20.3 & 3.0 & -40.1 & 274.1 & $2.2 \times 10^{-4}$ & 0.03 \\ 

\end{tabular}
      }
      \end{minipage}
    }
\end{table*}

\subsection{Candidate list search}
\label{CL}

\begin{figure*}[ht]
\centering
\includegraphics[width=0.48\linewidth]{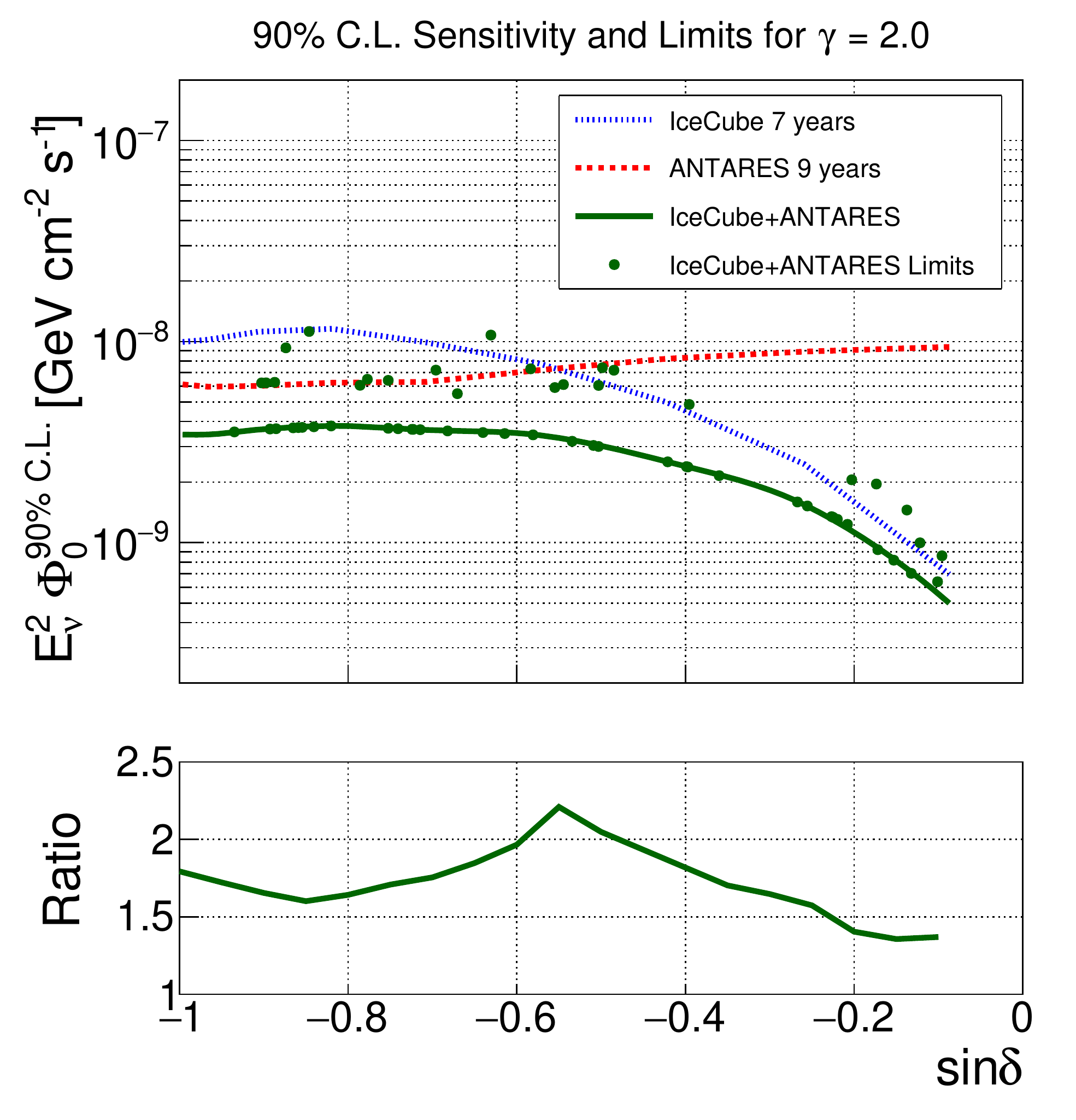}
\includegraphics[width=0.48\linewidth]{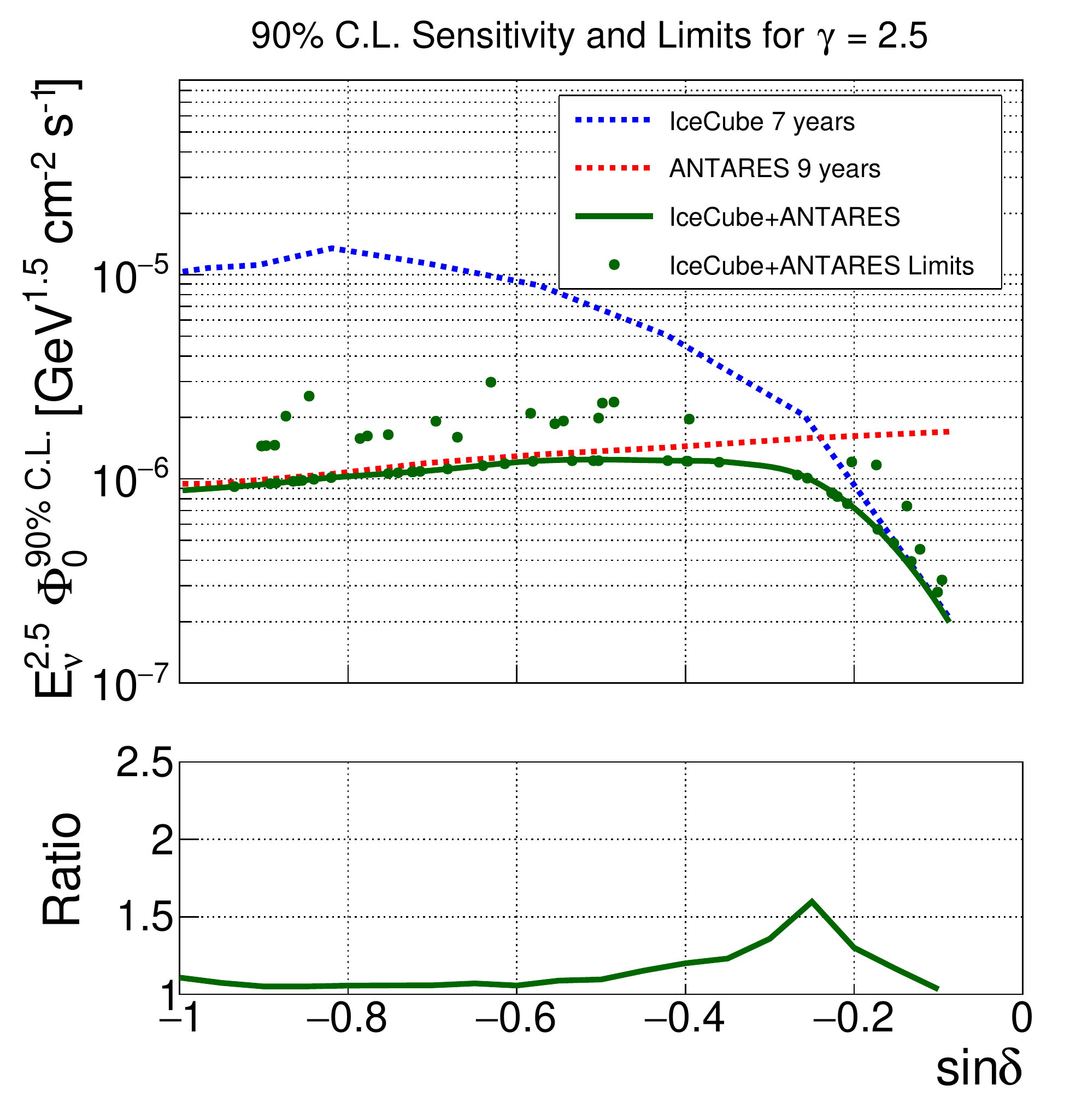}      
\caption{Top: upper limits at 90\% C.L. on the one-flavour neutrino flux normalization from the analysed candidates (green dots) reported in Table \ref{tab:LimitsFix2} as a function of the source declination. An unbroken $E_{\nu}^{-\gamma}$ neutrino spectrum is assumed, with $\gamma = 2.0$ (left) and $\gamma = 2.5$ (right). The green line indicates the sensitivity of the combined analysis. The dashed curves indicate the sensitivities for the IceCube (blue) and ANTARES (red) individual analyses. Bottom: ratio between the best individual sensitivity and the combined sensitivity as a function of the source declination for the spectral indices $\gamma = 2.0$ (left) and $\gamma = 2.5$ (right).}
\label{fig:LimitsCL}
\end{figure*}

In this study, the location of 57 astrophysical objects is investigated to look for point-like emission of high-energy neutrinos. The candidates are sources of high-energy $\gamma$-rays and belong to different object classes. The analysed candidates correspond to all the sources in the Southern Sky considered in the candidate list search performed by the ANTARES~\cite{lastPSant} and the IceCube~\cite{lastPSic} Collaborations. Here, only the number of signal events and the spectral index are left as free parameters in the likelihood maximisation as the direction of the selected sources is known. The list of the astronomical candidates is shown in Table~\ref{tab:LimitsFix2}, together with their equatorial coordinates, fitted number of signal events, fitted spectral index, pre-trial p-value, and 90\% C.L. upper limits on the one-flavour neutrino flux normalization for an $E_{\nu}^{-2.0}$ and an $E_{\nu}^{-2.5}$ spectrum. Figure \ref{fig:LimitsCL} shows the 90\% C.L. upper limits as a function of the source declination together with the median sensitivity.

\setlength{\tabcolsep}{.3em}
\begin{table*}[p]
    \caption{\footnotesize List of astrophysical objects analysed in the candidate list search. Reported are the source's name, equatorial coordinates, best-fit values of the free parameters, pre-trial p-value and 90\% C.L. upper limits on the one-flavour neutrino flux normalization for an $E_{\nu}^{-2.0}$ spectrum, $\Phi^{90\% \ \rm C.L.}_{0, \gamma = 2.0}$ (in units of $10^{-9} \ \rm{GeV^{-1} cm^{-2} s^{-1}}$), and for an $E_{\nu}^{-2.5}$ spectrum, $\Phi^{90\% \ \rm C.L.}_{0, \gamma = 2.5}$ (in units of $10^{-6} \ \rm{GeV^{-1} cm^{-2} s^{-1}}$). Dashes (-) in the fitted number of source events, spectral index and pre-trial p-value indicate sources with null observations ($\hat{n}_{\rm s} = 0.001$).
  \medskip}
    \label{tab:LimitsFix2}
    \resizebox{\linewidth}{!}{
        \begin{minipage}{0.9\textwidth} 
        \centering
 {\def\arraystretch{0.8}
    \footnotesize
    \begin{tabular}{crrrcccrr}
            Name & $\delta [\si{\degree}]$   &  $sin\delta$ &  $\alpha [\si{\degree}]$    & $\hat{n}_{\rm s}$  & $\hat{\gamma}$ & p-value & $\Phi^{90\% \ \rm C.L.}_{0, \gamma = 2.0}$ & $\Phi^{90\% \ \rm C.L.}_{0, \gamma = 2.5}$ \\
\hline
& & & & & & & \\
LHA120-N-157B & -69.16 & -0.93 & 84.43 & - & - & - & 3.6 & 0.9 \\ 
HESSJ1356-645 & -64.50 & -0.90 & 209.00 & 1.2 & 3.1 & 0.18 & 6.2 & 1.4 \\ 
PSRB1259-63 & -63.83 & -0.90 & 195.70 & 1.3 & 4.0 & 0.19 & 6.2 & 1.5 \\ 
HESSJ1303-631 & -63.20 & -0.89 & 195.74 & - & - & - & 3.7 & 0.9 \\ 
RCW86 & -62.48 & -0.89 & 220.68 & 1.0 & 1.6 & 0.20 & 6.3 & 1.5 \\ 
HESSJ1507-622 & -62.34 & -0.89 & 226.72 & - & - & - & 3.7 & 1.0 \\ 
HESSJ1458-608 & -60.88 & -0.87 & 224.54 & 3.7 & 3.6 & 0.036 & 9.3 & 2.0 \\ 
ESO139-G12 & -59.94 & -0.87 & 264.41 & - & - & - & 3.7 & 1.0 \\ 
MSH15-52 & -59.16 & -0.86 & 228.53 & - & - & - & 3.7 & 1.0 \\ 
HESSJ1503-582 & -58.74 & -0.85 & 226.46 & - & - & - & 3.7 & 1.0 \\ 
HESSJ1023-575 & -57.76 & -0.85 & 155.83 & 6.4 & 3.5 & 0.0079 & 11.2 & 2.5 \\ 
CirX-1 & -57.17 & -0.84 & 230.17 & - & - & - & 3.8 & 1.0 \\ 
SNRG327.1-01.1 & -55.08 & -0.82 & 238.65 & - & - & - & 3.8 & 1.0 \\ 
HESSJ1614-518 & -51.82 & -0.79 & 243.58 & 1.6 & 4.0 & 0.21 & 6.1 & 1.6 \\ 
HESSJ1616-508 & -50.97 & -0.78 & 243.97 & 2.0 & 2.0 & 0.18 & 6.5 & 1.6 \\ 
PKS2005-489 & -48.82 & -0.75 & 302.37 & 0.4 & 2.9 & 0.18 & 6.4 & 1.6 \\ 
GX339-4 & -48.79 & -0.75 & 255.70 & - & - & - & 3.7 & 1.1 \\ 
HESSJ1632-478 & -47.82 & -0.74 & 248.04 & - & - & - & 3.7 & 1.1 \\ 
RXJ0852.0-4622 & -46.37 & -0.72 & 133.00 & - & - & - & 3.7 & 1.1 \\ 
HESSJ1641-463 & -46.30 & -0.72 & 250.26 & - & - & - & 3.7 & 1.1 \\ 
VelaX & -45.60 & -0.71 & 128.75 & - & - & - & 3.6 & 1.1 \\ 
PKS0537-441 & -44.08 & -0.70 & 84.71 & 1.6 & 2.2 & 0.098 & 7.2 & 1.9 \\ 
CentaurusA & -43.02 & -0.68 & 201.36 & - & - & - & 3.6 & 1.1 \\ 
PKS1424-418 & -42.10 & -0.67 & 216.98 & 0.6 & 2.3 & 0.24 & 5.5 & 1.6 \\ 
RXJ1713.7-3946 & -39.75 & -0.64 & 258.25 & - & - & - & 3.5 & 1.2 \\ 
PKS1440-389 & -39.14 & -0.63 & 220.99 & 3.0 & 2.4 & 0.0085 & 10.8 & 3.0 \\ 
PKS0426-380 & -37.93 & -0.61 & 67.17 & - & - & - & 3.5 & 1.2 \\ 
PKS1454-354 & -35.67 & -0.58 & 224.36 & 3.9 & 2.1 & 0.089 & 7.3 & 2.1 \\ 
PKS0625-35 & -35.49 & -0.58 & 96.78 & - & - & - & 3.4 & 1.2 \\ 
TXS1714-336 & -33.70 & -0.55 & 259.40 & 1.2 & 2.3 & 0.17 & 5.9 & 1.9 \\ 
SwiftJ1656.3-3302 & -33.04 & -0.55 & 254.07 & 2.8 & 2.1 & 0.15 & 6.1 & 1.9 \\ 
PKS0548-322 & -32.27 & -0.53 & 87.67 & - & - & - & 3.2 & 1.2 \\ 
H2356-309 & -30.63 & -0.51 & 359.78 & - & - & - & 3.0 & 1.2 \\ 
PKS2155-304 & -30.22 & -0.50 & 329.72 & - & - & - & 3.0 & 1.2 \\ 
HESSJ1741-302 & -30.20 & -0.50 & 265.25 & 1.0 & 2.9 & 0.12 & 6.0 & 2.0 \\ 
PKS1622-297 & -29.90 & -0.50 & 246.50 & 4.4 & 1.9 & 0.048 & 7.4 & 2.4 \\ 
Sagittarius A* & -29.01 & -0.48 & 266.42 & 2.9 & 2.1 & 0.06 & 7.2 & 2.4 \\ 
Terzan5 & -24.90 & -0.42 & 266.95 & - & - & - & 2.5 & 1.2 \\ 
1ES1101-232 & -23.49 & -0.40 & 165.91 & - & - & - & 2.4 & 1.2 \\ 
PKS0454-234 & -23.43 & -0.40 & 74.27 & - & - & - & 2.4 & 1.2 \\ 
W28 & -23.34 & -0.40 & 270.43 & 1.7 & 2.5 & 0.094 & 4.9 & 2.0 \\ 
PKS1830-211 & -21.07 & -0.36 & 278.42 & - & - & - & 2.2 & 1.2 \\ 
NRG015.4+00.1 & -15.47 & -0.27 & 274.52 & - & - & - & 1.6 & 1.0 \\ 
LS5039 & -14.83 & -0.26 & 276.56 & - & - & - & 1.5 & 1.0 \\ 
QSO1730-130 & -13.10 & -0.23 & 263.30 & - & - & - & 1.3 & 0.9 \\ 
HESSJ1826-130 & -13.01 & -0.23 & 276.51 & - & - & - & 1.3 & 0.8 \\ 
HESSJ1813-126 & -12.68 & -0.22 & 273.34 & - & - & - & 1.3 & 0.8 \\ 
1ES0347-121 & -11.99 & -0.21 & 57.35 & - & - & - & 1.2 & 0.8 \\ 
PKS0727-11 & -11.70 & -0.20 & 112.58 & 2.5 & 2.7 & 0.13 & 2.1 & 1.2 \\ 
HESSJ1828-099 & -9.99 & -0.17 & 277.24 & 2.4 & 2.9 & 0.079 & 2.0 & 1.2 \\ 
HESSJ1831-098 & -9.90 & -0.17 & 277.85 & - & - & - & 0.9 & 0.6 \\ 
HESSJ1834-087 & -8.76 & -0.15 & 278.69 & - & - & - & 0.8 & 0.5 \\ 
PKS1406-076 & -7.90 & -0.14 & 212.20 & 6.8 & 2.7 & 0.11 & 1.5 & 0.7 \\ 
QSO2022-077 & -7.60 & -0.13 & 306.40 & - & - & - & 0.7 & 0.4 \\ 
HESSJ1837-069 & -6.95 & -0.12 & 279.41 & 2.5 & 3.4 & 0.24 & 1.0 & 0.5 \\ 
2HWCJ1309-054 & -5.49 & -0.10 & 197.31 & 9.1 & 3.2 & 0.051 & 0.9 & 0.3 \\ 
3C279 & -5.79 & -0.10 & 194.05 & 2.5 & 2.2 & 0.28 & 0.6 & 0.3 \\  
\ 

  \end{tabular}
      }
      \end{minipage}
      
    }
  
\end{table*}

The most significant source of the list is HESSJ1023-575, a TeV $\gamma$-ray source coincident with the young stellar cluster Westerlund 2~\cite{HESSsource}, with a pre-trial p-value of 0.79\%. 
A total of 6.4 signal events and a spectral index of 3.5 are fitted for the cluster at the position of HESSJ1023-575. The trial-corrected significance of the cluster is 42\%, corresponding to 0.2$\sigma$ in the one-sided convention. The second and third most significant sources are PKS1440-389 and HESSJ1458-608 with a p-value of 0.85\% and 3.6\%, respectively.

\subsection{Sagittarius A*}
\label{SG}

\begin{figure*}[ht]
\centering
\includegraphics[width=0.5\linewidth]{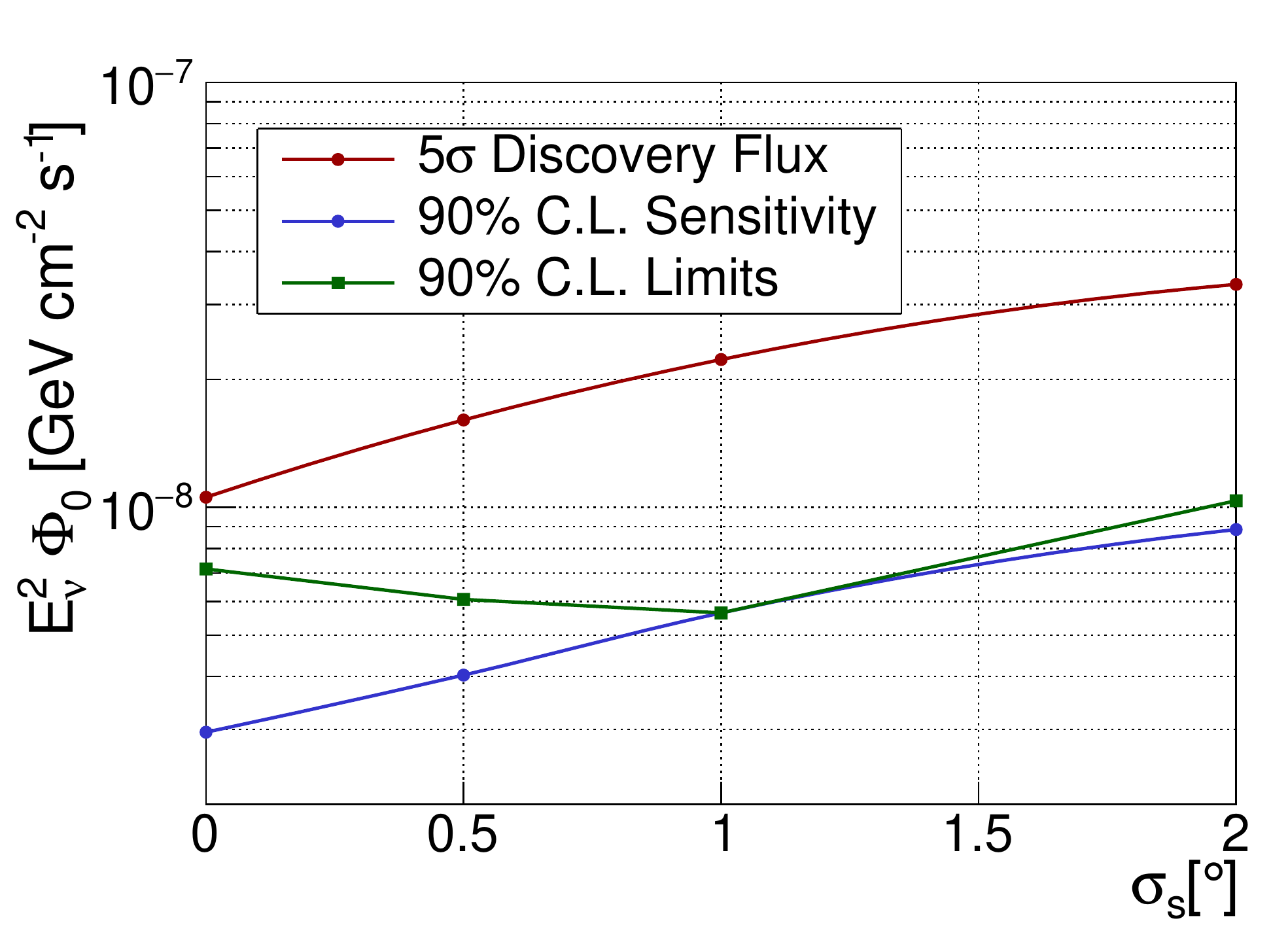}
\caption{Discovery flux (red dots), median sensitivity (blue dots) and 90\% C.L. upper limits (green squares) for the search at the location of Sagittarius A*, assuming an $E_{\nu}^{-2.0}$ spectrum, as a function of the angular extension $\sigma_{\rm s}$.}
\label{fig:LimitsSG}
\end{figure*}

Sagittarius A*, the SMBH located at the centre of our Galaxy, $(\alpha, \delta) = (266.42^{\circ}, -29.01^{\circ})$, is a candidate source of particular interest. Indeed, the surroundings of this kind of black holes are highly plausible acceleration sites of very-high-energy cosmic rays, and therefore, possible sources of cosmic neutrinos~\cite{SMBH1, SMBH2}. The high density of candidate objects and the presence of molecular clouds around the Galactic Centre makes the detection of an extended source more likely than the detection of a point-like source. For these reasons, a search for astrophysical neutrinos from Sagittarius A* and nearby objects is carried out by testing the point-like ($\sigma_{\rm s} = 0.0^\circ$) and extended source ($\sigma_{\rm s} = 0.5^\circ$, $1.0^\circ$, $2.0^\circ$) hypotheses.
The values of the best-fit parameters at the investigated location for the various tested source extensions are presented in Table~\ref{tab:SGtable} together with the observed p-value. 
The largest excess above the background is found for a point-like source hypothesis, with best-fit $\hat{n}_{\rm s} = 2.9$ and $\hat{\gamma} = 2.1$, and a significance of $6\%$ (1.6$\sigma$ in the one-sided sigma convention). The $90\%$ C.L. upper limits on the one-flavour neutrino flux normalization as a function of the source extension are shown in Figure \ref{fig:LimitsSG} together with the median sensitivity and the discovery flux.

\setlength{\tabcolsep}{.3em}
\begin{table*}[ht!]
    \label{tab:SGtable}
    \caption{Values of the best-fit parameters (number of signal events, $\hat{n}_{\rm s}$, and spectral index, $\hat{\gamma}$) and p-value for the search at the location of Sagittarius A* for different hypotheses of source extension $\sigma_{\rm s}$. Dashes (-) in the fitted number of source events, spectral index and p-value indicate cases of null observations ($\hat{n}_{\rm s} = 0.001$).    \medskip}
    \resizebox{\linewidth}{!}{
        \begin{minipage}{0.8\textwidth} 
        \centering
 {\def\arraystretch{1.}
    \footnotesize
    \begin{tabular}{c|c|c|c}
   \hline
            $\sigma_{\rm s} [\si{\degree}]$ & $\hat{n}_{\rm s}$ & $\hat{\gamma}$ &  p-value \\
\hline

0.0 & 2.9 & 2.1 & 0.06 \\ 
0.5 & 0.6 & 2.0 & 0.26 \\ 
1.0 & - & - & - \\ 
2.0 & 0.3 & 3.8 & 0.40 \\ 

\end{tabular}
      }
      \end{minipage}
    }
\end{table*}

\subsection{RXJ 1713.7-3946}
\label{RXJ}

Supernova remnants (SNRs) are the prime candidates for the acceleration of Galactic cosmic rays, and hence potential sources of astrophysical neutrinos. In the past years a large number of Galactic SNRs have been identified by $\gamma$-ray telescopes~\cite{tevcat}. Some of the observed $\gamma$-ray spectra extend up to tens of TeV suggesting that these objects are accelerators of high-energy particles. The observation of neutrinos from these sources would be an unambiguous indication of hadronic acceleration. The shell-type SNR RXJ~1713.7–3946, at equatorial coordinates $(\alpha, \delta) = (258.25^{\circ}, -39.75^{\circ})$, is the brightest object of this kind in the TeV $\gamma$-ray sky and represents a particularly interesting target to the search for cosmic neutrinos~\cite{RXJ1, RXJ_KAPPES, RXJ_VISSANI}. In this analysis, two different models are considered for the neutrino emission: that proposed by Kappes et al.~\cite{RXJ_KAPPES}, in the following indicated as RXJ 1713.7-3946 (1), and the one recently introduced for KM3NeT neutrino source search estimations~\cite{KM3NeTPS} and based on the methods described by Vissani et al.~\cite{RXJ_VISSANI, RXJ_VISSANI2, RXJ_VISSANI3}, hereafter referred to as RXJ~1713.7-3946 (2).
Both models describe a neutrino spectrum of the form of:

\begin{align} \label{eq:specRXJ}
\Phi_{ \nu_{f} + \bar{\nu}_{f}} = \Phi_0  \ \left( \frac{E_\nu}{1\,{\rm TeV}} \right)^{-\Gamma} \exp [-(E_{\nu}/E_{\rm{cut}})^{\beta}],
\end{align} 

\noindent where $E_{\nu}$ is the neutrino energy and the values of the neutrino spectrum parameters $\Phi_0$, $\Gamma$, $E_{\rm{cut}}$, and $\beta$ are listed in Table~\ref{tab:RXJtable}. A Gaussian extension with $\sigma_{\rm s} = 0.6^\circ$ is assumed for the source as reported by the $\gamma$-ray analysis performed by the H.E.S.S. Collaboration~\cite{HESS}. 

No significant evidence of astrophysical neutrinos from the direction of the SNR is observed for either of the considered spectra. The fitted number of signal events and the p-value observed at the source position are presented in Table~\ref{tab:RXJtable} for each spectrum hypothesis, together with the 90\% C.L. sensitivity and upper limit, both expressed as ratio with the theoretical  source flux.

\setlength{\tabcolsep}{.3em}
\begin{table*}[ht!]
    \label{tab:RXJtable}
    \caption{List of considered neutrino emission models for the search at the location of RXJ~1713.7–3946 and respective results. For each model, the values of the neutrino spectrum parameters, $\Phi_0$ (in units of $10^{-11} \rm{TeV}^{-1} \rm{cm}^{-2} \rm{s}^{-1}$), $\Gamma$, $E_{\rm{cut}}$ (in units of TeV) and $\beta$, entering Eq.~\ref{eq:specRXJ} are provided. The last four columns show the results in terms of best-fit number of signal events, $\hat{n}_{\rm s}$, p-value, ratio of the sensitivity to the assumed source flux, $\Phi^{90\% \rm{C.L.}}_{\rm sens}/\Phi_0$, and ratio of the upper limit to the assumed source flux, $\Phi^{90\% \rm{C.L.}}_{\rm UL}/\Phi_0$.      
    \medskip}
    \resizebox{\linewidth}{!}{
        \begin{minipage}{0.8\textwidth} 
        \centering
 {\def\arraystretch{1.4}
    \footnotesize
    \begin{tabular}{c|c|c|c|c|c|c|c|c}
   \hline
            Spectrum & $\Phi_0$ & $\Gamma$ & $E_{\rm{cut}} $ & $\beta$ & $\hat{n}_{\rm s}$ & p-value & $\Phi^{90\% \rm{C.L.}}_{\rm sens}/\Phi_0$ & $\Phi^{90\% \rm{C.L.}}_{\rm UL}/\Phi_0$ \\
\hline

RXJ 1713.7-3946 (1) & $1.55$ & 1.72 & 1.35 & 0.5 & 0.3 & 0.40 & 10.7 & 13.2\\
RXJ 1713.7-3946 (2) & $0.89$ & 2.06 & 8.04 & 1 & 0.3 & 0.41 & 9.7 & 11.7\\

\end{tabular}
      }
      \end{minipage}
    }
\end{table*}

\section{Conclusions}
\label{Conc}

A combined search for neutrino sources in the Southern Sky using data from the ANTARES and IceCube telescopes was presented. Neither significant point-like nor extended neutrino emission over the background expectation was found. 

The largest excess over the whole Southern Sky, with a post-trial significance of $18\%$, was found at equatorial coordinates ($\hat{\alpha}, \hat{\delta}) = (213.2\si{\degree}, -40.8\si{\degree})$, for a point-like source hypothesis. When limiting the search to the GC region, the most significant cluster was found at equatorial coordinates ($\hat{\alpha}, \hat{\delta}) = (274.1\si{\degree}, -40.1\si{\degree})$, with a post-trial significance of $3\%$, for a source extension of $2.0\si{\degree}$. Upper limits on the neutrino flux from 57 astrophysical candidate sources were presented. The most significant source candidate is HESSJ1023-575 with a post-trial significance of $42\%$. The upper limits on the flux from HESSJ1023-575 were set to $1.1\times10^{-8}(2.5\times10^{-6})$ in units of $\rm{GeV^{-1} cm^{-2} s^{-1}}$ for an unbroken power-law spectrum with spectral index $\gamma = 2.0$ ($\gamma = 2.5$). Sagittarius A* was tested as a point-like source and as an extended source. The largest excess over the background was observed at an angular extension of $0.0\si{\degree}$ with a significance of $6\%$. Finally, the location of the SNR RXJ~1713.7-3946 was investigated assuming two proposed neutrino emission models and a source extension of $0.6\si{\degree}$. As no significant evidence of cosmic neutrinos was observed, upper limits were derived.

This analysis shows the strong potential to search for neutrino sources in the Southern Sky using the joint data sets of the ANTARES and IceCube telescopes. The combination of the two detectors, which differ in size and location, allows for an improvement of up to a factor $\sim$2 in the sensitivity in different regions of the Southern Sky, depending on the energy spectrum of the source. For a soft spectral index, the contribution of high energy neutrinos is suppressed and ANTARES dominates in most of the Southern Sky. The complementarity of the two detectors is mostly effective for a harder spectral index as all the samples provide a significant contribution. For an $E_{\nu}^{-2.0}$ spectrum a considerable gain in the sensitivity to point-like sources is achieved in all the Southern Sky and in a larger scale in the region close to the Galactic Centre.

\section*{Acknowledgements}
\parindent=0em
The authors of the ANTARES collaboration acknowledge the financial support of the funding agencies:
Centre National de la Recherche Scientifique (CNRS), Commissariat \`a
l'\'ener\-gie atomique et aux \'energies alternatives (CEA),
Commission Europ\'eenne (FEDER fund and Marie Curie Program),
Institut Universitaire de France (IUF), IdEx program and UnivEarthS
Labex program at Sorbonne Paris Cit\'e (ANR-10-LABX-0023 and
ANR-11-IDEX-0005-02), Labex OCEVU (ANR-11-LABX-0060) and the
A*MIDEX project (ANR-11-IDEX-0001-02),
R\'egion \^Ile-de-France (DIM-ACAV), R\'egion
Alsace (contrat CPER), R\'egion Provence-Alpes-C\^ote d'Azur,
D\'e\-par\-tement du Var and Ville de La
Seyne-sur-Mer, France;
Bundesministerium f\"ur Bildung und Forschung
(BMBF), Germany; 
Istituto Nazionale di Fisica Nucleare (INFN), Italy;
Nederlandse organisatie voor Wetenschappelijk Onderzoek (NWO), the Netherlands;
Council of the President of the Russian Federation for young
scientists and leading scientific schools supporting grants, Russia;
Executive Unit for Financing Higher Education, Research, Development and Innovation (UEFISCDI), Romania;
Ministerio de Ciencia, Innovaci\'{o}n, Investigaci\'{o}n y Universidades (MCIU): Programa Estatal de Generaci\'{o}n de Conocimiento (refs. PGC2018-096663-B-C41, -A-C42, -B-C43, -B-C44) (MCIU/FEDER), Severo Ochoa Centre of Excellence and MultiDark Consolider (MCIU), Junta de Andaluc\'{i}a (ref. SOMM17/6104/UGR), 
Generalitat Valenciana: Grisol\'{i}a (ref. GRISOLIA/2018/119), Spain; 
Ministry of Higher Education, Scientific Research and Professional Training, Morocco.
We also acknowledge the technical support of Ifremer, AIM and Foselev Marine
for the sea operation and the CC-IN2P3 for the computing facilities.

The authors of the IceCube collaboration acknowledge the support from the following agencies and institutions: USA – U.S. National Science Foundation-Office of Polar Programs, U.S. National Science Foundation-Physics Division, Wisconsin Alumni Research Foundation, Center for High Throughput Computing (CHTC) at the University of Wisconsin-Madison, Open Science Grid (OSG), Extreme Science and Engineering Discovery Environment (XSEDE), U.S. Department of Energy-National Energy Research Scientific Computing Center, Particle astrophysics research computing center at the University of Maryland, Institute for Cyber-Enabled Research at Michigan State University, and Astroparticle physics computational facility at Marquette University; Belgium – Funds for Scientific Research (FRS-FNRS and FWO), FWO Odysseus and Big Science programmes, and Belgian Federal Science Policy Office (Belspo); Germany – Bundesministerium für Bildung und Forschung (BMBF), Deutsche Forschungsgemeinschaft (DFG), Helmholtz Alliance for Astroparticle Physics (HAP), Initiative and Networking Fund of the Helmholtz Association, Deutsches Elektronen Synchrotron (DESY), and High Performance Computing cluster of the RWTH Aachen; Sweden – Swedish Research Council, Swedish Polar Research Secretariat, Swedish National Infrastructure for Computing (SNIC), and Knut and Alice Wallenberg Foundation; Australia – Australian Research Council; Canada – Natural Sciences and Engineering Research Council of Canada, Calcul Québec, Compute Ontario, Canada Foundation for Innovation, WestGrid, and Compute Canada; Denmark – Villum Fonden, Danish National Research Foundation (DNRF), Carlsberg Foundation; New Zealand – Marsden Fund; Japan – Japan Society for Promotion of Science (JSPS) and Institute for Global Prominent Research (IGPR) of Chiba University; Korea – National Research Foundation of Korea (NRF); Switzerland – Swiss National Science Foundation (SNSF); United Kingdom – Department of Physics, University of Oxford.

\bibliographystyle{utphys}

\providecommand{\href}[2]{#2}\begingroup\raggedright\endgroup

\end{document}